\documentstyle[psfig]{aa}

\begin{document}
\thesaurus { }

\title{ Interaction rate at $z\sim1$ } 

\author{Vladimir P. Reshetnikov}

\offprints{resh@astro.spbu.ru}   

\institute{Astronomical Institute of St.Petersburg State
University, 198904 St.Petersburg, Russia}  

\date{Received June 8; accepted October 8, 1999}

\maketitle
\markboth{V.P.Reshetnikov}{Interaction rate at $z\sim1$}

\begin{abstract}
We found 25 galaxies with probable tidal tails in the Hubble
Deep Fields North and South at $z$=0.5--1.5. General characteristics of the
selected tidal features are very close to characteristics
of tidal tails of local interacting galaxies. Using objects
with $z$=0.5--1.0, we found that volume density of
galaxies with tidal tails changes with $z$ as (1\,+\,$z$)$^{4\pm1}$.
Therefore, we estimated the rate of close encounters
between the galaxies of comparable mass leading to the formation
of extended tidal structures. If this rate reflects the merger rate,
our data support a steeply increasing merger rate at $z\sim$1 and
consistent with zero curvature universe.

\end{abstract}

\section{Introduction}

The redshift dependence of the interaction and merger rate is an
important test of the current models for the formation and evolution
of galaxies. It is now well established that
galaxy interactions play a major role in galaxy formation and evolution
(e.g. Schweizer 1998, Combes 1999 for recent reviews).
Toomre (1977) demonstrated that the universe's higher density in the
past (\,$\propto$(1+$z$)$^3$\,) suggests a higher past merger
rate, increasing back in time as $t^{-5/3}$ (with time $t$) if the binding
energies of binary galaxies had a flat distribution.
If the galaxy merger rate is parametrized in the power-law
form $\propto(1+z)^m$, then the exponent has been found to be
$m=2.5$ from Toomre's (1977) approach (assuming $\Omega=1$).
Statistics of close galaxy pairs from
faint-galaxy redshift surveys (e.g. Yee \& Ellingson 1995,
Le Fevre et al. 1999)
and morphological studies of distant galaxies support a large value of
the exponent $m$ for $z\leq1$.
For instance, Abraham (1998) concluded that current
best estimates for the merger rate are consistent with $m\approx3$.
Preliminary studies of distant peculiar objects representing
distinct results of interactions/mergers (collisional ring galaxies,
polar-ring galaxies, mergers) also support $m\geq3$ (Lavery et al. 1996,
Reshetnikov 1997, Remijan et al. 1998, Le Fevre et al. 1999),
although statistics are still insufficient.
Many other surveys, including IRAS faint sources,
or quasars, have also revealed a high power-law (e.g. Warren et al. 1994,
Springel \& White 1998). However, some recent works have suggested
a moderate ($m\leq2$) (e.g. Neuschaefer et al. 1997, Wu \& Keel 1998)
or intermediate ($m\sim2-3$) (Burkey et al. 1994, Im et al. 1999) density 
evolution of merging systems with $z$.

From analytical formulation of merging histories (e.g.
Carlberg 1990, 1991; Lacey \& Cole 1993), it is possible to relate the
dark haloes merger rate to the parameters of the universe (average
density $\Omega$, cosmological constant $\Lambda$). The merging rates for
visible galaxies should follow, although the link is presently not well known
(Carlberg 1990, Toth \& Ostriker 1992). Theoretical models based on
Press-Schechter formalism (Carlberg 1990, 1991) predict a redshift 
evolution of the merger rate with
$m~\propto~\Omega^{0.42}(1-\Lambda)^{-0.11}$ (the exponents must be
somewhat changed if the average halo mass decreases with $z$ --
Carlberg et al. 1994). This conclusion is confirmed by numerical simulations
within the CDM scenario -- $m=4.2$ ($\Omega=1$) and $m=2.5$ ($\Omega=0.3$) 
for $z\leq1$ (Governato et al. 1997).

Tidal tails originate in close encounters of disk galaxies
(e.g. Toomre \& Toomre 1972).
The purpose of this note is to show that statistics of galaxies with
extended tidal tails (tidal bridges have, on average, fainter surface
brightnesses -- Schombert et al. 1990) is a useful tool to study evolution 
of interaction rate to $z\sim1$. The simulations
by Hibbard \& Vacca (1997) -- they showed that extended tidal features
remain readily visible in the long exposures typical of the Hubble Deep
Fields out to $z\sim1$ -- is the theoretical base for our work.
We found that current statistics of such objects in
the North and South Hubble Deep Fields (HDF-N and HDF-S correspondingly)
leads to $m\approx4$. (Preliminary results based on the HDF-N only are
presented in Reshetnikov 1999 -- Paper I.)

\section{Sample of galaxies}

We used the deepest currently available deep fields (HDF-N -- Williams et al. 
1996 and HDF-S -- Williams et al. 1999) to search galaxies with extended 
tidal tails.
From detailed examination of the fields in the F814W filter (hereinafter
referred to as $I$), we selected more than 70 tailed objects.
Careful analysis of their images in combination with the redshift data
enabled us to distinguish 25 galaxies with $z=0.5-1.5$ (12 objects in
the HDF-N and 13 in the HDF-S). Galaxies with tidal tails in the HDF-N
are described in detail in Paper I. Here we present the data for
the HDF-S objects (Fig.1). 
(Our statistics of galaxies with tidal structures are in general agreement
with van den Bergh et al. (1996) data on morphology of galaxies in
the HDF-N. van den Bergh et al. classified 20 galaxies with
$21<I<25$ as objects with probable tidal distortions in the HDF-N.)

\begin{figure}
\psfig{file=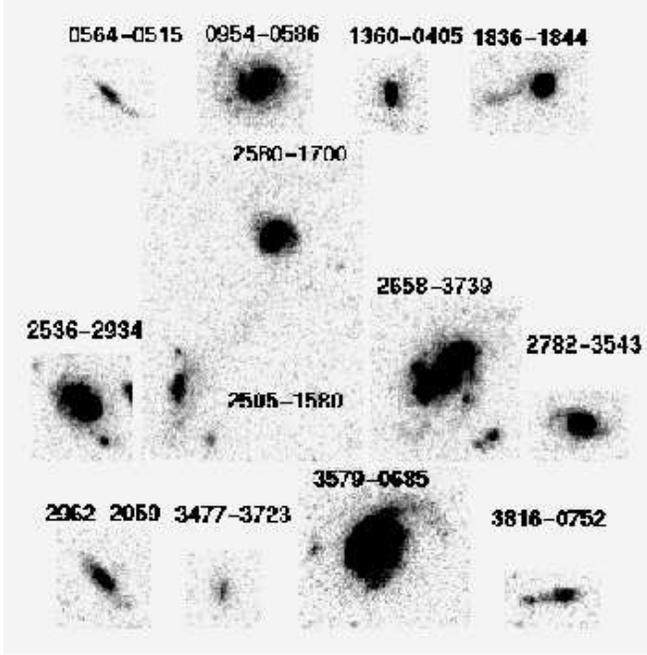,height=8.7cm}
\caption{ The $I$-band images of the HDF-S galaxies with tidal tails.
The size of the panel is 19''$\times$19''.}
\end{figure}

General characteristics of the galaxies are summarized in Table 1.
The columns of the table are: galaxy identification, $I$ band apparent
magnitude, photometric redshift (there are no published spectroscopic
redshifts for the sample galaxies). All the data are taken from 
the web site of the HDF-S group at SUNY, Stony Brook (Chen et al. 1998).
In the fourth column we present the absolute magnitude in the rest-frame
$B$ band calculated according to Lilly et al. (1995) as: \\
$M_B$=$I$--5\,log($D_L$/10\,pc)+2.5\,log(1+$z$)+($B-I_{z}$)+0.17, \\
where
\begin{equation}
D_L~=~\frac{c}{H_0\,q_0^2}[q_0z\,+\,(q_0\,-\,1)(\sqrt{2q_0z\,+\,1}\,-\,1)] 
\end{equation}
-- luminosity distance ($\Lambda=0$),
$H_0$ -- the Hubble constant (75 km/s/Mpc), $q_0$ -- deceleration parameter
($q_0$=0.05), $B-I_z$ -- $k$-correction color (we used correction for
Sbc galaxy), and term 0.17 translates AB magnitudes into standard $B$.

\begin{table*}
\caption[1]{Characteristics of galaxies in the HDF-S}
\begin{tabular*}{12cm}{@{\extracolsep{\fill}}|c|c|c|c|c|c|}
\hline 
        &           &              &       &          &     \\
Name    & $I_{\rm F814W}$ & $z_{ph}$ & $M_B$ & $\mu_I$(tail) & 
$\mu_B$(tail) \\
     &        &  &      &     &      \\
\hline
SB-WF-0564-0515 & 25.27      & 1.01    & -17.4  &25.4 & 23.1 \\
\hline
SB-WF-0954-0586 & 22.63      & 1.29    & -21.1  &24.9 & 21.6 \\
\hline
SB-WF-1360-0405 & 24.26      & 1.02    & -18.4  &25.1 & 22.8 \\
\hline	
SB-WF-1836-1844 & 22.85      & 0.91    & -19.4  &24.9 & 23.0 \\
\hline
SB-WF-2505-1580 & 24.04      & 1.37    & -20.1  &25.4 & 21.6 \\
\hline
SB-WF-2536-2934 & 22.68      & 1.00    & -20.0  &24.9 & 22.6 \\
\hline
SB-WF-2580-1700 & 22.62      & 1.27    & -21.0  &26.2 & 22.9 \\
\hline
SB-WF-2658-3739 & 21.78      & 0.47    & -18.7  &25.7 & 25.1 \\
\hline
SB-WF-2782-3543 & 23.42      & 1.08    & -19.5  &24.7 & 22.2 \\
\hline
SB-WF-2962-2059 & 24.20      & 0.51    & -16.5  &25.1 & 24.4 \\
\hline
SB-WF-3477-3723 & 25.79      & 0.95    & -16.7  &25.8 & 23.7 \\
\hline
SB-WF-3579-0685 & 21.33      & 0.58    & -19.7  &24.9 & 24.0 \\
\hline
SB-WF-3816-0752 & 24.54      & 0.69    & -16.9  &24.4 & 23.1 \\
\hline
\end{tabular*}
\end{table*}

In Table 2 we compare mean characteristics of the tailed galaxies
in two deep fields. As one can see, both samples are consistent within
quoted errors.

\begin{table}
\caption[2]{Mean characteristics of interacting galaxies}
\begin{tabular}{|c|c|c|}
\hline 
             &           &                 \\
Parameter    &  HDF North & HDF South \\
             &           &            \\
\hline
             &           &            \\
Number of galaxies (all) & 12  &  13 \\
$z=$0.5--1.0 &     7     &   7       \\
$z=$1.0--1.5 &     5     &   6       \\
$<z>$        &  0.94$\pm$0.30 & 0.93$\pm$0.28 \\
$<M_{B}>$    & --18.0$\pm$1.3 & --18.9$\pm$1.5 \\
$<\mu_{B}>$(tail) & 23.7$\pm$1.0 & 23.1$\pm$1.0   \\
d$_{tail}$/d$_{galaxy}$& 1.4$\pm$0.7 & 1.2$\pm$0.5 \\
             &           &     \\
\hline	     
\end{tabular}
\end{table}

\section{Characteristics of tidal tails}

To be sure that our selected galaxies possess true tidal tails,
we performed photometric measurements in the $I$ passband using
circular apertures centered on the brightest regions of the suspected tails.
For the measurements, we retrieved the HDF-S images (version 1)
from the ST~ScI web site and processed them in the ESO-MIDAS environment.
The results of these measurements are summarized in Table 1
(column 5). The observed surface brightness of the tails has been
converted to a rest-frame $B$ by applying the cosmological dimming
term and a $k$-correction color term:
$\mu(B)~=~\mu(I)$~--~2.5\,log(1+$z$)$^3$~+~($B-I_{z}$)~+~0.17
(Lilly et al. 1998).
General photometric characteristics of the local tidal tails are
close to those for late-type spiral galaxies (Sb-Sc) (Schombert et al. 1990, 
Reshetnikov 1998) and we used color term for Sbc galaxy 
(Lilly et al. 1995) in our calculations. The results are presented in
the last column of Table 1 and in Fig.2.

The mean rest-frame surface brightness of the tidal structures in the joint 
(HDF-N plus HDF-S) sample is $<\mu_{B}>$(tail) = 23.4$\pm$1.1, in full 
agreement with our results for the local sample of interacting galaxies 
(obtained by analogous manner) -- $<\mu_{B}>$(tail) = 23.8$\pm$0.8
(Reshetnikov 1998).

Fig.2 (top) presents the observed distribution of the $<\mu(I)>$
values of the suspected HDF tails (dashed line) in comparison
with the distribution for local objects in the $B$ passband.
The bottom part of the figure shows converted to a rest-frame, $B$
distribution for HDF tails.  It is evident that tails of distant
galaxies demonstrate a distribution of $<\mu(B)>$ values close
to that for local interacting galaxies. Fig.2 illustrates clearly
the influence of observational selection on the recognition of
tidal structures -- we are able
to detect relatively faint tails among the galaxies with $z=0.5-1.0$
but among $z=1.0-1.5$ objects
we can see only very bright tails. Therefore, our sample of
galaxies with extended tidal tails is sufficiently incomplete for
$z\geq1$. Thus, objects with $z=0.5-1.0$ will give a more reasonable
estimation of $m$ in comparison with the total sample.

\begin{figure}
\psfig{file=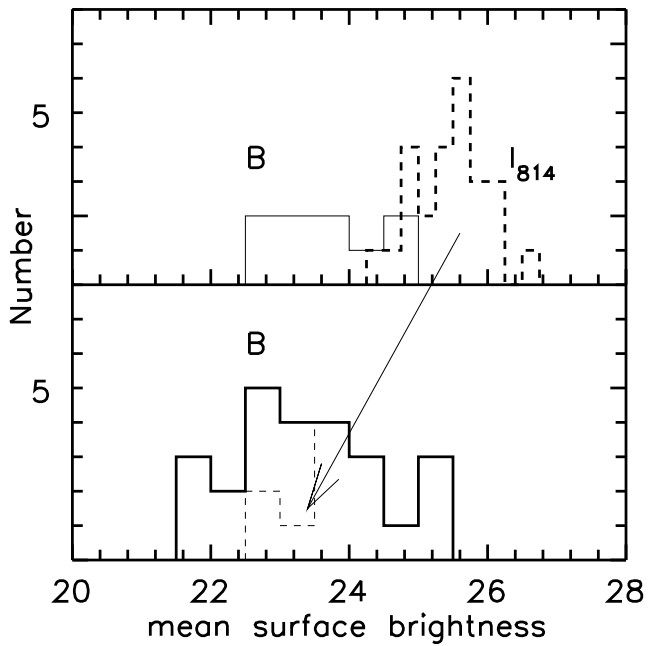,width=9.2cm}
\caption{The distribution of surface brightnesses of tidal tails.
{\it Top}: solid line -- distribution for local galaxies in the $B$ band
according to Reshetnikov (1997), dashed line -- distribution for
objects in the HDF-N and HDF-S in the $I$ filter. {\it Bottom}: distribution
of the $B$ band rest-frame surface brightnesses for HDF tails
(calculated from observed $I$ band distribution). Dashed line shows
galaxies with $z=0.5-1.0$.}
\end{figure}

\section{Density evolution}

The resemblance of morphological and photometric characteristics
of suspected tidal tails of the HDFs galaxies with local objects
allows us to use them to measure possible change with $z$ of volume density
of galaxies with tails (and, therefore, the rate of close encounters
leading to the formation of extended tails). 

The co-moving volume element in solid angle $d\Omega$ and redshift
interval $dz$ is
\begin{equation}
d{\rm V}~=~\frac{c}{H_0}\,(1~+~z)^{-2}\,\frac{D_L^2}{E(z)}\,d\Omega\,dz,
\end{equation}
where $D_L$ -- photometric distance (eq.(1)), and
$E(z)~=~$(1\,+\,$z$)$\sqrt{2q_0z\,+\,1}$ for $\Lambda=0$ (e.g. Peebles 1993).
The increase of the space density of galaxies with tidal tails we
take in standard power-law form:
\begin{equation}
n(z)~=~n_0~(1\,+\,z)^{m}, 
\end{equation}
where $n_0=n(z=0)$ -- local volume density of such galaxies. 
By integrating equations (2) and (3) we can find the expected
number of objects within solid angle $d\Omega$ and in required
range of $z$.

\subsection{Local density of galaxies with tidal tails}

We suppose that at the current epoch interactions and mergers
accompanied by tail formation are almost entirely between bound pairs
of galaxies (e.g. Toomre 1977). So we adopt that frequency of tidal
tails among single objects (mergers) and in groups, is significantly
lower than in pairs. 

According to Karachentsev (1987), the relative frequency of galaxies
with tails among the members of binary systems is
94/974=0.10$\pm$0.01. The fraction of paired galaxies in the local
universe is not well determined. Various strategies give results
between 5\% and 15\%. For instance, local pairing fraction is
7\%$\pm$1\% according to Burkey et al. (1994),
6\%-10\% (Keel \& van Soest 1992), 14\%$\pm$2\% (Lawrence et al. 1989).
The most intensive studies lead to 12\%$\pm$2\% (Karachentsev 1987)
and 10\% (Xu \& Sulentic 1991, Soares et al. 1995). Moreover,
Xu \& Sulentic (1991)
found that the fraction of pairs is approximately constant (10\%)
over the luminosity range $-22<M_B<-16$ (see also Soares et al. 1995).
Thus, we can adopt the value of 10\%$\pm$5\% as a reasonable estimate
of the local fraction of binary galaxies. Therefore, the relative
fraction of galaxies with tidal tails at $z=0$ is
0.1$\times$0.1=0.01$\pm$0.005.

To find total density of galaxies in the nearby part of the universe
($z\leq0.05$), we considered the galaxy luminosity function (LF) according to
Marzke et al. (1998). The adopted Schechter function parameters of the
LF are: $M_B^*$=--20.05, $\phi^*$=5.4$\times$10$^{-3}$ Mpc$^{-3}$ and
$\alpha$=--1.12 ($H_0=$75 km/s/Mpc). By integrating LF from $M_B=-15.4$
to --21.1 (the range of absolute luminosities of galaxies with tails
in the HDF-N and HDF-S), we found that total volume density of galaxies
is equal to 0.026 Mpc$^{-3}$.
Thus, $n_0=$0.01$\times$0.026=(2.6$\pm$1.3)$\times$10$^{-4}$\,Mpc$^{-3}$.

The total angular area within which we searched tailed
galaxies in two HDFs is 10.4 arcmin$^2$ or 8.8$\times$10$^{-7}$ sr.

\subsection{Exponent $m$ from tidal structures}

Varying exponent $m$, we can estimate the expected number of galaxies
with tidal features in the HDFs. In Fig.3 we present the results
of calculations for two redshift intervals: 0.5--1.5 (total sample)
and 0.5--1.0 (adopted cosmology is $\Lambda=0$, $q_0=0.05$ or
$\Omega=2q_0=0.1$ and $H_0$=75 km/s/Mpc). As one can see, the total
sample (25 objects) leads to $m=2.6$. But this value must be considered
as a low limit only due to strong underestimation of tidal tails at
$z\geq1$ (sect.3). For the galaxies with $z=0.5-1.0$ ($N$=14)
we obtain $m=4.0$. Assuming Poisson error of $N$ ($\sqrt{N}$=3.7),
we have $m=4.0^{+0.4}_{-0.5}$. Adding 50\% uncertainty in
the local space density $n_0$, we have obtained a final estimation of $m$
as 4.0$^{+1.2}_{-0.9}$. (Let us note also that two
potential sources of errors -- underestimation of $n_0$ value
and omission of tailed galaxies in the HDFs -- bias value of $m$
in opposite directions and partially compensate each other.)

\begin{figure}
\psfig{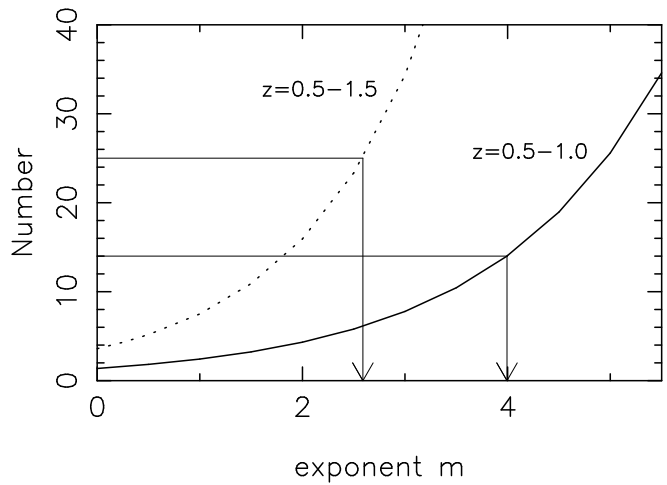}
\caption{The dependence of expected number of galaxies with tidal tails
in two HDFs on exponent $m$ for $z$=0.5--1.5 (dotted line) and
$z$=0.5--1.0 (solid line). Horisontal lines show observed quantities.}
\end{figure}

The value of $m$ depends on the adopted cosmological model. For
$\Lambda=0$, $\Omega=1$ we have $m=4.9$ ($z$=0.5--1.0). In our
calculations for the model with a cosmological constant and zero spatial
curvature ($\Omega_m$=0.3, $\Omega_{\Lambda}$=0.7,
$\Omega=\Omega_m+\Omega_{\Lambda}$=1) we used the analytical approximation
of the luminosity distance $D_L$ according to Pen (1999). In the
framework of that model we have obtained $m=3.6^{+1.2}_{-0.9}$.

To obtain more realistic error estimation, we must take into account
the possible luminosity evolution of galaxies with redshift. Unfortunately,
luminosity and surface brightness evolution of peculiar and interacting
galaxies is poorly constrained at present (e.g. Roche et al. 1998).
Moreover, Simard et al. (1999) claim that an apparent systematic increase
in disk mean surface brightness to $z\sim1$ for bright ($M_B<-19$)
spiral galaxies is due to selection effects. Nevertheless, assuming
that interacting galaxies undergo luminosity evolution $\Delta\,M_B=1^m$
between $z=0$ and 1, we estimated that the value of $m$ must be decreased by
$\Delta\,m\approx0.5$: $m=3.5$ for $\Omega=0.1,~\Lambda=0$ and
$m=4.4$ for $\Omega=1,~\Lambda=0$.

\section{Discussion and conclusions}

On the basis of analysis of the HDF-N and HDF-S images we selected
25 galaxies with probable tidal tails with $z$=0.5--1.5. Integral
photometric characteristics of the suspected tails are close to
that for local interacting galaxies. Considering the subsample of
tailed galaxies with $z$=0.5-1.0 (14 objects), we estimated that
co-moving volume density of such galaxies changes approximately
as (1+$z$)$^4$. (Hence the volume density of tailed galaxies at
$z=1$ is $n(z=1)=4\times10^{-3}~{\rm Mpc}^{-3}$ for $q_0=0.05$.)
Inclusion in the sample of the galaxies with
tidal bridges does not noticeably change the value of the exponent (Paper I).
Therefore, we estimated the change of the rate of close encounters
leading to the formation of extended tails. If this rate reflects
the merger rate, we have obtained evidence of a steeply increasing
merger rate at $z\sim1$. (Our result is related to field galaxies.
The evolution in clusters might even be stronger than in the field.
For instance, van Dokkum et al. 1999 found $m=6\pm2$ for the merger
fraction in rich clusters of galaxies.)

\begin{figure}
\psfig{file=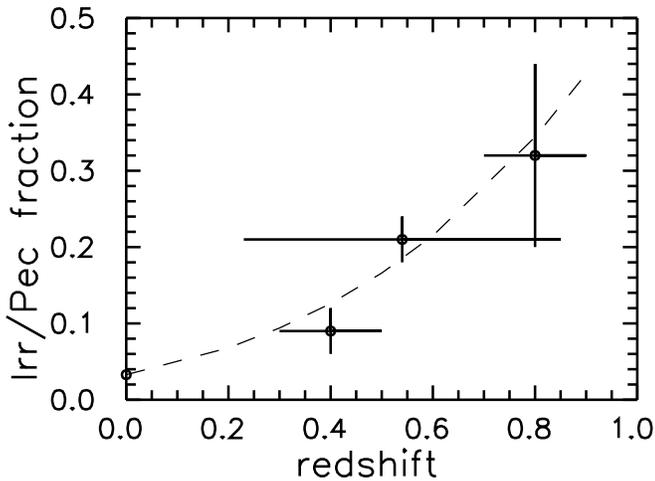,width=9.2cm}
\caption{The dependence of the relative fraction of irregular and
peculiar galaxies on redshift. The data for $z$=0.4 and 0.8 are
from Brinchmann et al. (1998); for $z$=0.54 are from Roche et al. (1998);
the local value ($z$=0) is estimated
from Marzke et al. (1998) LF. Dashed line -- (1+z)$^4$ relation.}
\end{figure}

How does our estimation of $m$ agree with values obtained by other methods?
The recent surveys of the evolution of galaxy pairs with $z$
are consistent with $m\approx3$ (see references in Abraham 1998).
Evolution of the rate of interactions according to our data
is characterized by close (within quoted errors) value of $m$.
Direct analysis of the morphology of distant galaxies at $z\leq1$
suggests a significant increase of the fraction of irregular and peculiar
systems with redshift (Fig.4). If interactions and mergers are responsible
for the observed asymmetries of galaxies (e.g. Conselice \& Bershady 1998),
this increase can reflect the increase of the interaction rate with $z$.
As one can see in Fig.4, relative fraction of Irr/Pec galaxies changes
in accordance with $m=4$. Naim et al.'s (1997) result (35\%$\pm$15\% of
peculiar galaxies down to $I=24.0$) agrees with $m=4$ also.
Many other observational surveys and numerical works indicate
a large ($\geq$3) exponent $m$ (Sect.1). Comparison with predictions
of analytical and numerical works shows that current observational
estimates of the merger rate favor a zero curvature ($\Omega=1$) universe
(e.g. Carlberg 1991, Governato et al. 1997).

Our results indicate that further detailed statistics of galaxies with
tidal structures will be a powerful tool to quantify the interaction and
merging rates evolution.

\acknowledgements{I would like to thank an anonymous referee for useful
comments. I acknowledge support from the Russian Foundation for Basic 
Research (98-02-18178) and from the ``Integration'' programme ($N$~578).}


\begin{thebibliography}{}
\bibitem{} Abraham R.G., 1998, in ``Galaxy interactions at low and high
redshifts'', Barnes J.E., Sanders D.B. eds. 1998. Kluwer, Dordrecht, 11
\bibitem{} Brinchmann J., Abraham R., Schade D. et al., 1998, ApJ 499, 112
\bibitem{} Burkey J.M., Keel W.C., Windhorst R.A., Franklin B.E., 1994,
ApJ 429, L13
\bibitem{} Carlberg R.G., 1990, ApJ 359, L1
\bibitem{} Carlberg R.G., 1991, ApJ 375, 429
\bibitem{} Carlberg R.G., Pritchet C.J., Infante L., 1994, ApJ 435, 540
\bibitem{} Chen H.-W., Fernandez-Soto A., Lanzetta K.M. et al., 1998,
astro-ph/9812339
\bibitem{} Combes F., 1999, in ``Building galaxies: from the
primordial universe to the present'', Hammer F., Thuan T.X.,
Cayatte V., Guiderdoni B., Tran Thanh Van J. eds. Frontieres (astro-ph/9904133)
\bibitem{} Conselice C.J., Bershady M.A., 1998, astro-ph/9812299
\bibitem{} Hibbard J.E., Vacca W.D., 1997, AJ 114, 1741
\bibitem{} Im M., Griffiths R.E., Naim A. et al., 1999, ApJ 510, 82
\bibitem{} Governato F., Garnder J.P., Stadel J. et al., 1997,
ApJ, astro-ph/9710140
\bibitem{} Karachentsev I.D., 1987, Binary galaxies. Nauka, Moscow
\bibitem{} Keel W.C., van Soest E.T.M., 1992, A\&AS 94, 553
\bibitem{} Lacey C., Cole S., 1993, MNRAS 262, 627
\bibitem{} Lavery R.J., Seitzer P., Suntzeff B. et al., 1996, ApJ 467, L1
\bibitem{} Lawrence A., Rowan-Robinson M., Leech K. et al., 1989,
MNRAS 240, 329
\bibitem{} Le Fevre O., Abraham R., Lilly S.J. et al., 1999, MNRAS,
astro-ph/9909211
\bibitem{} Lilly S.J., Tresse L., Hammer F. et al., 1995, ApJ 455, 108
\bibitem{} Lilly S., Schade D., Ellis R. et al., 1998, ApJ 500, 75
\bibitem{} Marzke R.O., da Costa L.N., Geller M.J., 1998, ApJ 503, 617
\bibitem{} Naim A., Ratnatunga K.U., Griffiths R.E., 1997, ApJ 476, 510
\bibitem{} Neuschaefer L.W., Im M., Ratnatunga K.U. et al., 1997,
ApJ 480, 59
\bibitem{} Peebles P.J.E., 1993, Principles of Physical Cosmology.
Princeton University Press, Princeton
\bibitem{} Pen U.-L., 1999, ApJS 120, 49
\bibitem{} Remijan A.J., Lavery R.J., Reed M.D., 1998, BAAS 193, 305
\bibitem{} Reshetnikov V.P., 1997, A\&A 321, 749
\bibitem{} Reshetnikov V.P., 1998, Pis'ma v AZh 24, 189 (Engl. transl.
in Astron. Letters 24, 153)
\bibitem{} Reshetnikov V.P., 1999, Pis'ma v AZh, in press (Paper I)
\bibitem{} Roche N., Ratnatunga K., Griffiths R.E. et al., 1998,
MNRAS 293, 157
\bibitem{} Schombert J.M., Wallin J.F., Struck-Marcell C., 1990, AJ 99, 497
\bibitem{} Schweizer F., 1998, in ``Galaxies: interactions and induced star
formation'': lecture notes 1996/Saas Fee Advanced Course 26,
Friedli D., Martinet L., Pfenniger D. eds. Springer, 105
\bibitem{} Simard L., Koo D.C., Faber S.M. et al., 1999, ApJ 519, 563
\bibitem{} Soares D.S.L., de Souza R.E., de Carvalho R.R.,
Couto da Silva T.C., 1995, A\&AS 110, 371
\bibitem{} Springel V., White S.D.M., 1998, MNRAS 298, 143
\bibitem{} Toomre A., 1977, in ``The evolution of galaxies and stellar
populations'', Tinsley B.M., Larson R.B. eds. Yale Univ. Obs., New Haven, 401
\bibitem{} Toomre A., Toomre J., 1972, ApJ 178, 623
\bibitem{} Toth G., Ostriker J.P., 1992, ApJ 389, 5
\bibitem{} van den Bergh S., Abraham R.G., Ellis R.S. et al., 1996,
AJ 112, 359
\bibitem{} van Dokkum P.G., Franx M., Fabricant D. et al., 1999, 
ApJL, accepted (astro-ph/9905394)
\bibitem{} Warren S.J., Hewett P.C., Osmer P.S., 1994, ApJ 421, 412
\bibitem{} Williams R.E., Blacker B., Dickinson M. et al., 1996,
AJ 112, 1335
\bibitem{} Williams R.E. et al., 1999, AJ, to be submitted
\bibitem{} Wu W., Keel W.C., 1998, AJ 116, 1513
\bibitem{} Xu C., Sulentic J.W., 1991, ApJ 374, 407
\bibitem{} Yee H.K.C., Ellingson E., 1995, ApJ 445, 37

\end{thebibliography}
\end{document}